\documentclass{aa}

\usepackage{graphicx}
\usepackage{multicol}
\usepackage{subcaption}
\usepackage{booktabs}
\usepackage[normalem]{ulem}
\usepackage{wrapfig}
\usepackage{balance}
\usepackage{amssymb}
\usepackage{flushend}
\usepackage{float}
\usepackage{xcolor}
\newcommand{\orcid}[1]{\href{https://orcid.org/#1}{\textcolor[HTML]{A6CE39}{\aiOrcid}}}
\usepackage{orcidlink}
\usepackage{url}
\usepackage{blindtext}
\usepackage{placeins}
\usepackage{txfonts}
\definecolor{darkgoldenrod}{rgb}{0.72, 0.53, 0.04}
\definecolor{dca}{rgb}{0.64, 0.0, 0.0}
\definecolor{pp}{rgb}{0.7, 0.0, 0.9}

\usepackage{soul}
\usepackage{hyperref}

\begin{document}

   \title{Spin measurement of 4U 1543--47 with \textit{Insight}-HXMT and \textit{NICER} from its 2021 outburst}
   \subtitle{A test of accretion disk models at high luminosities}

   \author{E. S. Yorgancioglu
           \inst{1}\fnmsep\thanks{yorgancioglu@astro.uni-tuebingen.de} \orcidlink{0000-0002-8442-9458}
            \and
            Q. C. Bu\inst{1}\fnmsep\thanks{bu@astro.uni-tuebingen.de} \orcidlink{0000-0001-5238-3988}
           \and
           A. Santangelo\inst{1} \orcidlink{0000-0003-4187-9560}
           \and
            L. Tao\inst{2,3} \orcidlink{0000-0002-2705-4338}
            \and 
            S. W. Davis\inst{4} \orcidlink{0000-0001-7488-4468}
            \and A. Vahdat\inst{1} \orcidlink{0000-0002-4026-5885}
            \and L. D. Kong\inst{1} \orcidlink{0000-0003-3188-9079}
            \and S. Piraino\inst{1} \orcidlink{0000-0003-0122-6899}
            \and M. Zhou\inst{1} \orcidlink{0000-0001-8250-3338}
            \and S. N. Zhang\inst{2,3}
          \orcidlink{0000-0001-5586-1017}
            }
            
   \institute{Institut f{\"u}r Astronomie und Astrophysik, Sand 1, 72076 T{\"u}bingen, Germany
   \and
   Key Laboratory of Particle Astrophysics, Institute of High Energy Physics, Chinese Academy of Sciences, Beijing 100049, People’s Republic of China
   \and 
   University of Chinese Academy of Sciences, Chinese Academy of Sciences, Beijing 100049, People’s Republic of China
   \and Department of Astronomy, University of Virginia, Charlottesville, VA 22904, USA}

   \date{Received --- ; accepted ---;}

 
  \abstract
   {4U 1543--47 is one of a handful of known black hole candidates located in the Milky Way galaxy. It underwent a very luminous outburst in 2021, reaching a peak intensity of  $\sim$9 Crab, as observed by the Monitor of All-sky Image (MAXI), and exceeding twice its Eddington luminosity.   }
   { The unprecedented bright outburst of 4U 1543--47 provides a unique opportunity to test the behavior of accretion disk models at high luminosities and accretion rates. In addition, we explore the possibility of constraining the spin of the source at high accretion rates, given that the previous spin measurements of 4U 1543--47 are largely inconsistent with each other. 
       }
   {We measure the spectral evolution of the source throughout its outburst as observed by \textit{Insight}-HXMT, and compare the behavior of both the thin-disk model \texttt{kerrbb2} and the slim disk model \texttt{slimbh} up to the Eddington limit for two different values of disk $\alpha$-viscosity. In addition, given the behavior of these two models, we identify two "golden" epochs in which it is most suitable to measure the spin with the continuum fitting (CF) method. }
   {We find evidence of a disk state transition from a thicker slim disk to a thin disk occurring around $1 L_{\rm{Edd}}$ from fits to the luminosity-temperature (LT) relation. We obtain consistent and constant spin measurements from both \texttt{slimbh} and \texttt{kerrbb2} as the luminosity varies towards the Eddington limit, implying the recovery of thin-disk solutions above the traditional thin-disk criterion of 30\% $L_{\mathrm{Edd}}$. We constrain the spin to $a_{*} = 0.65^{+0.14}_{-0.24}$, assuming an $\alpha$-viscosity = 0.01 from both \textit{Insight}-HXMT and \textit{NICER} observations from the above-mentioned "golden" epochs where the condition of the disk being truncated at the innermost stable circular orbit (ISCO) is most closely met. }
   {}


\maketitle
%

\section{INTRODUCTION}

    Black holes (BHs) are fully defined by their mass, angular momentum, and charge, as stipulated by the no-hair theorem of general relativity (GR). However, in practice, a BH would inevitably undergo a discharge process to its environment, causing it to become neutral. Consequently, it is reasonable to assume that the space-time around BHs is well-represented by the Kerr Metric \citep{kerr1963gravitational}. The angular momentum of a black hole is characterized by the dimensionless spin parameter, $a_{*} \equiv cJ/GM^2 $, where $M$ is the BH mass, $J$ is the angular momentum, $G$ is Newton's gravitational constant, c is the speed of light, and  $-1 \leq  a_{*} \leq 1$. \
   
    There are two principal methods to extract the spin of a BH in accreting systems: the continuum fitting method \citep{Zhang_1997, Li2005},  which requires accurate measurements of distance $D$, inclination $i$, and BH mass $M$, and the X-ray reflection method \citep{iwa1997, mil2002}, which is weakly dependent on inclination angle. More recently, various methods of spin extraction have been proposed, including the relativistic precession model (RPM;  \citealp{stella1999khz}), and the X-ray polarization method \citep{dovvciak2008thermal}. Given sufficiently sensitive instruments, spin measurements in the nonaccreting regime are also possible via gravitational waves signals (see review by \citealt{Reynolds_2021}).
   
   The thermal continuum method, or continuum fitting (CF), relies on the fundamental assumption that the accretion disk extends up to the innermost stable circular orbit (ISCO), and is truncated there. As the ISCO radius is a monotonic function of the spin, fitting the thermal continuum by identifying $R_{\mathrm{in}}$ from the temperature maximum of soft X-rays would allow us to measure $a_{*}$.  Typically, the outburst is screened for the soft state dominated by thermal emission. Thin-disk solutions, which model a geometrically thin and optically thick disk, are typically realized when the disk luminosity $l_{\mathrm{Disk}} \leq 30\%  L_{\mathrm{Edd}}$ \citep{nov1973, sha1973}.  
   
   Among black hole transients (BHTs), 4U 1543--47 has been the subject of relatively extensive investigation due to its peculiar outburst and outflow history. The recurrent  X-ray binary was discovered by the \textit{Uhuru} satellite in 1971, and was observed in subsequent outbursts in 1983, 1992 and 2002 \citep{mat1972, kit1984, har1992, par2004}. There have been several attempts to measure its spin, three of which used continuum fitting. \cite{sha2006} first measured a spin of 0.80 $\pm$ 0.05. Thereafter, \cite{mil2009} and  \cite{mor2014} measured spin values of 0.3 $\pm$ 0.1 and  $0.43^{+0.22}_{-0.31}$, respectively. All of these works use the fiducial values of $M = 9.4 \pm 2.0 M_{\odot} $ \citep{orosz2003inventory}, $i = 20.7^{\circ} \pm 1.5$ \citep{orosz2003inventory}, and $D = 7.5 \pm 0.5$ kpc \citep{jonker2004distances}, with the exception of \cite{mil2009} who used an inclination of $ i = {{32^{\circ}}^{+3}_{-4}}$ obtained based on their constraints to the iron line. The latest spin measurement by \cite{Dong2020} found a moderately high black hole spin value of $0.67^{+0.15}_{-0.08}$ via X-ray reflection spectroscopy. 
   
\begin{figure}[h]
    \centering
    \includegraphics[width = 0.50\textwidth]{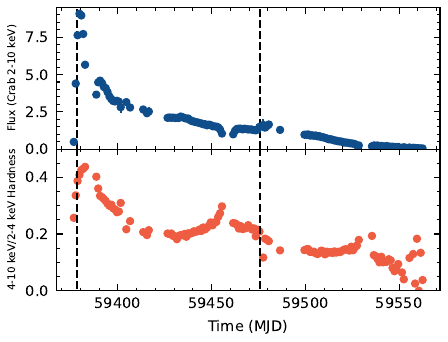}
    \caption{Temporal evolution of the light-curve \& hardness ratio of 4U 1543--47 measured by all-sky monitor MAXI. The dashed black lines denote the start and end dates of \textit{Insight}-HXMT monitoring.}
    \label{fig:lc_hr}
\end{figure}

    4U 1543--47 underwent a bright outburst in 2021, reaching a peak flux of $\sim$ 9 Crab, as observed by the Monitor of All-sky X-ray Image (MAXI) (see Figure~\ref{fig:lc_hr}). The outburst was the brightest ever observed by \textit{NICER} to date \citep{Connors20224U}.  Throughout the entirety of both \textit{Insight}-HXMT and \textit{NICER} observations, the source remains above 30\% of its Eddington luminosity, and as such, no observations formally satisfy the traditional luminosity constraint employed when using continuum fitting. In addition to the thin-disk model, we therefore adopt the model {\ttfamily slimbh} in XSPEC \citep{skadowski2009slim, Straub2011}, which is a generalization of the Shakura-Sunyaev thin-disk model and can account for luminous, optically thick disks, or "slim" disks \citep{Abramowiczslim} at high luminosities. 
   
   In this paper,  we report the performance of both thin- and slim-disk models at high luminosities extending up to 1 $L_{\rm{Edd}}$, using \textit{Insight}-HXMT observations of the source from its 2021 outburst. Given the performance of the models, we explore the possibility to constrain the spin of 4U 1543--47 with continuum fitting at these relatively high accretion rates, using both \textit{Insight}-HXMT and \textit{NICER} datasets.  As in previous studies, we use fiducial values of $M$, $i$, and $D$ for our source. The paper is organized as follows. In section 2, we give an introduction to relativistic slim disks. In section 3, we provide details on the data reduction and selection. In section 4, we present our results, including the spectral evolution of the source, the behavior of  slim- and thin-disk models, and provide our own spin measurement. Sections 5 and 6 present our discussion and conclusions, respectively. Unless otherwise stated, our errors are always reported within 1$\sigma$.


\section{Slim-disk \textit{versus} Standard thin-disk models }

 In the standard thin-disk model of the accretion disk \citep{sha1973, nov1973}, energy advection is neglected and all the gravitational energy release is in the form of radiation. The  "slim-disk" describes a thicker accretion disk ($h/r  \gtrsim  0.03$) at high luminosities, and takes into account three additional effects not assumed to be present in thin disk solutions \citep{Abramowiczslim}. In particular, the slim-disk model takes into account (1) heat advection, which becomes more dominant at higher luminosities and modifies the flux profiles of radiation emitted in the inner disk region; (2) at higher luminosities approaching the Eddington limit, radiation pressure begins to play a more significant role, causing a deviation from Keplerian angular momentum, especially in the inner disk region.  The inner edge of the disk may also be much closer to the black hole than the ISCO radius due to heat advection.  \cite{Straub2011} presents the flux profiles for three different accretion rates for a moderately rotating black hole. At high accretion rates, advection causes emission from within the ISCO to become more dominant. Compared to the Novikov-Thorne (NT) thin-disk model, a significant portion of the heat generated by disk $\alpha$-viscosity is radiated closer to the BH horizon, and some heat may be advected into the horizon and thus never  emitted. The effects of advection are clearly demonstrated in Figure 1 of \cite{Straub2011};  
(3) the location of the photosphere, which may diverge from the equatorial plane at higher luminosities \citep{Sadowski2009, Straub2011}.

\def\arraystretch{1.1}
\begin{table*}[ht]
  \centering
    \begin{tabular}{lcccccc}
     \multicolumn{6}{l}{{\textbf{Table 1}. \textit{Insight}-HXMT `golden' observation log}} \\[0.2cm]
        \hline
          Observation                 & Date (MJD)             & LE Exposure (ks) &  ME Exposure (ks)                    & LE Rate (cts/s) & ME Rate (cts/s)              \\
        \hline
          P030402603501                   & 59461.3         &  2.4 & 3.1               &  1959 ± 1 &  192.5 ± 0.2          \\
          P030402603502               & 59461.4          &  2.1 & 2.2            &  1990 ± 1 & 189.6 ± 0.3            \\
          P030402603503     & 59461.6        & 1.4 & 1.7        &  1987 ± 12 & 181.4 ± 0.3         \\
          P030402603601     & 59464.4        & 0.9 &  2.6       &  2029 ± 15 & 196.9 ± 0.3         \\
          P030402603602     & 59464.6        & 1.9 & 1.6        &  2032 ± 10 & 171.4 ± 0.3         \\
          P030402603603     & 59464.7        & 1.5 &  1.6      &  2037 ± 12 & 169.9 ± 0.3         \\
          P030402603701     & 59466.5        & 0.7  &  1.8    &  1946 ± 16 & 174.1 ± 0.3        \\
          P030402603702     & 59466.6        & 1.4 & 2.1       & 1922  ± 12 & 159.9 ± 0.3         \\
          P030402603703     & 59466.7        & 1.3 & 1.2       &  1918 ± 12 & 154.9 ± 0.4         \\
          P030402603801     & 59468.2        & 1.6 & 2.5        &  1949 ± 11 & 192.9 ± 0.3         \\
          P030402603802     & 59468.3        & 1.4 & 2.2        &  1913 ± 12 & 177.5 ± 0.3         \\
          P030402603803     & 59468.5        & 1.8 & 2.1       &  1965 ± 10 & 172.3 ± 0.3         \\
          P030402604101     & 59476.2        & 1.4 &  3.6      &  1847 ± 12 & 176.9 ± 0.2         \\
        \hline
    \end{tabular}
     
    \caption*{ \textbf{Notes.} Energy bands for LE is 1--10 keV, while for ME is 10--30 keV. }
    \captionlistentry{Table}
  \label{tab:hxmtobs}
\end{table*}

\begin{table*}[ht]
  \centering
    \begin{tabular}{lcccc}
    \multicolumn{4}{l}{{\textbf{Table 2}. \textit{NICER} `golden' observation log}} \\[0.2cm]
        \hline
          Observation             & Date (MJD)             & Exposure (ks)               & Rate (cts/s)              \\
        \hline
        
          4202230162               & 59461.1          &  1.2       & 24330 $\pm$ 5             \\
          4202230165               & 59464.2          &  0.6    & 24070 $\pm$ 6             \\
          4202230166               & 59466.1          &  1.7     &  23650 $\pm$ 4             \\
          4202230167               &  59467.1          &  1.3    & 23460 $\pm$ 4             \\
          4202230168               & 59468.6          &  2.5      & 23810 $\pm$ 3             \\
       
          4202230172               & 59472.4          &  0.7     & 23250 $\pm$ 8             \\

        \hline
    \end{tabular}
    \caption*{\textbf{Notes.} All observations utilized 50 FPM, with energies between 0.2--12 keV. }
    \captionlistentry{Table}
  \label{tab:nicerlog}
\end{table*}

\section{Observation and data reduction}

 We use \textit{Insight}-HXMT observations (see Figure \ref{fig:lc_hr} for \textit{Insight}-HXMT's observation epoch) to measure the spectral evolution of the source, and test the performance of the relativistic slim- and thin-disk models \footnote{http://archive.hxmt.cn/proposal}. More specifically, we test the consistency of the spin estimates for both models over a luminosity range extending up to the Eddington limit. The performance and agreement between the models would serve as a nontrivial test of these models in attempting to measure spin at high accretion rates. This same procedure was employed by \cite{Straub2011}, who find that both relativistic slim-disk and thin-disk models suffer from a spin drop-off at luminosities above 30\% $L_{\rm{Edd}}$ when measured for LMC X-3. Measuring spin at high luminosities is associated with more uncertainties, and so it is therefore imperative to test both models to see if they give relatively consistent and constant spin values over a wide range of luminosities before providing constraints on the spin.

 In order to find a suitable epoch for spin extraction, that is where the edge of the accretion disk is truncated at the ISCO, we adopt the following criteria: (i) a low, stable inner disk radius epoch; and (ii) spectra dominated by thermal disk emission, where the proportion of thermal disk photons that scatter in the corona (the scattering fraction, $f_{\mathrm{sc}}$) does not exceed 10\%. It is important that the suitable epoch chosen for spin extraction has the lowest values of the normalization or inner disk radius relative to the other observations, when measured by the multi-color disk blackbody model (\texttt{diskbb} in XSPEC). Moreover, results by \cite{ste2009b} showed that when the disk emission accounted for over 75\% of the total luminosity ($f_{\mathrm{sc}}$ < 25\%), the spin measurements remained relatively constant. These selection criteria for CF were  successfully employed by  \cite{gou2009determination}, \cite{steiner2011spin}, and \cite{chen2016spin}, and can also be extended to other spectral states besides the high soft state (HSS). Taken together, these two criteria will mean that the assumption of the disk being truncated at the ISCO is most justified. Any observations vigorously meeting these conditions are designated as "golden" spectra. Because all observations  exceed the thin-disk luminosity limit, we use the slim-disk model {\ttfamily slimbh} to measure the spin for all "golden" observations and give a final spin result. Spectral fitting was carried out using XSPEC \textsc{v. 12.12.1} and HEASoft version \textsc{v6.30}.

\subsection{\textit{Insight}-HXMT}

\textit{Insight}-HXMT is China's first X-ray Astronomy satellite, and was launched on June 15, 2017. Its scientific payload consists of the low-energy (LE) X-ray telescope \citep{Chen_2020}, covering 1--15 keV (384 c$\mathrm{m^2}$), the medium-energy (ME) X-ray telescope \citep{cao_2020}, covering 8--35 keV (952 c$\mathrm{m^2}$), and the high-energy (HE) X-ray telescope \citep{Liu2020new}, covering 20--250 keV (5100 c$\mathrm{m^2}$). We extracted spectra using the software HXMTDAS \textsc{v2.05} \footnote{http://hxmten.ihep.ac.cn/SoftDoc/501.jhtml} and the pipeline prescribed by the official user guide. The latest calibration database was also used (CALDB \textsc{v2.06}). The following criteria were used for extraction: (1) An elevation angle $\geq$ 10°; (2) a geomagnetic cut-off rigidity of $\geq$ 8 GeV; (3) a pointing position offset of $\leq$ 0.05°; and (4) at least 300 s away from the South Atlantic Anomaly (SAA).  Out of 151 sub exposures, 108 survived this screening for Good Time Interval (GTI), all of which were used to measure the spectral evolution of the source. 
We only used data from the LE and ME telescopes, as they provide sufficient coverage of the relevant energy range for continuum fitting, and because of the very low photon count in HE. The backgrounds are estimated with the tools provided by the \emph{Insight}-HXMT team: LEBKGMAP \citep{liao2020backgroundLE} and MEBKGMAP \citep{guo2020background}, version 2.0.9 based on the current standard \emph{Insight}-HXMT background models, for LE and ME, respectively. We select a total of 13 exposures for the final spin analysis, which we refer to as the golden observations (see Table \ref{tab:hxmtobs}).

\subsection{NICER}

The Neutron Star Interior Composition Explorer (\textit{NICER}), a payload on the International Space Station (ISS), provides coverage of the soft X-ray range (0.2--12 keV). The X-ray timing instrument (XTI) of \textit{NICER} is composed of 56 identical and co-aligned cameras, each of which contains an X-ray concentrator (XRC, \citealt{Okajima}) and silicon drift detector (SDD) pairs. With a peak effective area of 1900 c$\mathrm{m^2}$ at 1.5 keV, it is well suited to thermal fitting, and hence spin measurements of BHs. We obtained cleaned event files by applying the standard calibration and filtering tool \textit{nicerl2}, and obtained the response files with NICERRMF and NICERARF tools with HEASOFT v 6.30. The background spectrum was computed using nibackgen3C50 tool. We select a total of six observations observed during the golden epoch to perform the spin extraction (see Table \ref{tab:nicerlog}).


\begin{table*}[ht]
  \centering
    \begin{tabular}{lcc}
    \multicolumn{2}{l}{{\textbf{Table 3}. Utilized Models}} \\[0.2cm]
        \hline
          Model         & Energy Bands              \\
        \hline
          M1: \texttt{constant$\times$TBabs$\times$Smedge$\times$(simpl$\circledast$diskbb)}                & 2-10 keV (LE), 10-30 keV (ME)         \\
          M2: \texttt{constant$\times$TBabs$\times$Smedge$\times$(Thcomp$\circledast$diskbb)}         & 2-10 keV (LE), 10-30 keV (ME)              \\
          M3: \texttt{constant$\times$TBabs$\times$Smedge$\times$(Thcomp$\circledast$kerrbb2)}         & 2-10 keV (LE), 10-30 keV (ME)              \\
           M4: \texttt{constant$\times$TBabs$\times$Smedge$\times$(Thcomp$\circledast$slimbh)}         & 2-8 keV (LE), 10-30 keV (ME)              \\
           M5: \texttt{TBabs$\times$edge$\times$Smedge$\times$slimbh}         & 2-9 keV (\textit{NICER})              \\
         
        \hline
    \end{tabular}
    \caption*{\textbf{Notes.} Spectral fitting with models M1-M2 were performed with \textit{Insight}-HXMT. Spin extraction with Models M3-M5 utilized both \textit{Insight}-HXMT (M3-M4) \& \textit{NICER} (M5).  Because all of the `golden' observations of \textit{Insight}-HXMT selected for spin measurement suffered from the photoelectric effect between 8--10 keV, we have restricted LE to 2--8 keV during the final spin estimate (M4). This was also done in the case of M1 \& M2 when measuring the spectral evolution on these observations. }
  \captionlistentry{Table}
  \label{tab:models}
\end{table*}

\vspace{4mm}

\section{Analysis and results}

\subsection{Accretion state evolution}

We use a total of five different models, which are labeled M1--M5 (see table  \ref{tab:models}). All used models account for interstellar absorption with the {\ttfamily TBabs} component in XSPEC, with  \citet{wil2000} abundances and \cite{ver1996} cross sections. We fix the column density, $N_{\text H}$, to 0.439$\times 10^{22} \mathrm{cm^{-2}}$ \citep{Connercol}. In addition, due to the presence of an excess at energies of between 6--8 keV, which is likely due to disk reflection, we use the model {\ttfamily Smedge} in XSPEC \citep{ebi1994} component in all models. The {\ttfamily SMEDGE} parameters are the absorption edge $E_{\mathrm{Edge}}$ between 7 and 9 keV, optical depth $\tau_{\mathrm{max}}$, photoelectric cross section (fixed to -2.67), and width (fixed at 7 keV). We study the spectral evolution of the source with the first two models, M1 and M2, both of which contain the multicolored absorbed blackbody component \texttt{diskbb} \citep{mit1984, Makshima1986}, accounting for the bulk of emission, but with differing Componization components: \texttt{Simpl} in M1 \citep{ste2009b},  and \texttt{ThComp} in M2 \citep{Zdziarski2020bMNRAS}. \texttt{ThComp} is a much more accurate representation of thermal comptonization compared to the simple phenomenological model \texttt{powerlaw}, taking into account the electron temperature $kT_{\mathrm{e}}$ and scattering fraction $f_{\mathrm{sc}}$. While not as comprehensive as \texttt{ThComp}, \texttt{simpl} is a semi-phenomenological model that nonetheless includes a parameter for the scattering fraction $f_{\mathrm{sc}}$, which is important for finding a suitable epoch for spin extraction. Both \texttt{ThComp} and \texttt{simpl} are convolution models that take in a thermal component as their seed spectrum.   These two models are well fitted, with an average reduced $\chi^2$ of 1.11 for M1 and 1.07 for M2. 

\begin{figure*}[h]
    \centering
    \includegraphics[width = 0.9\linewidth]{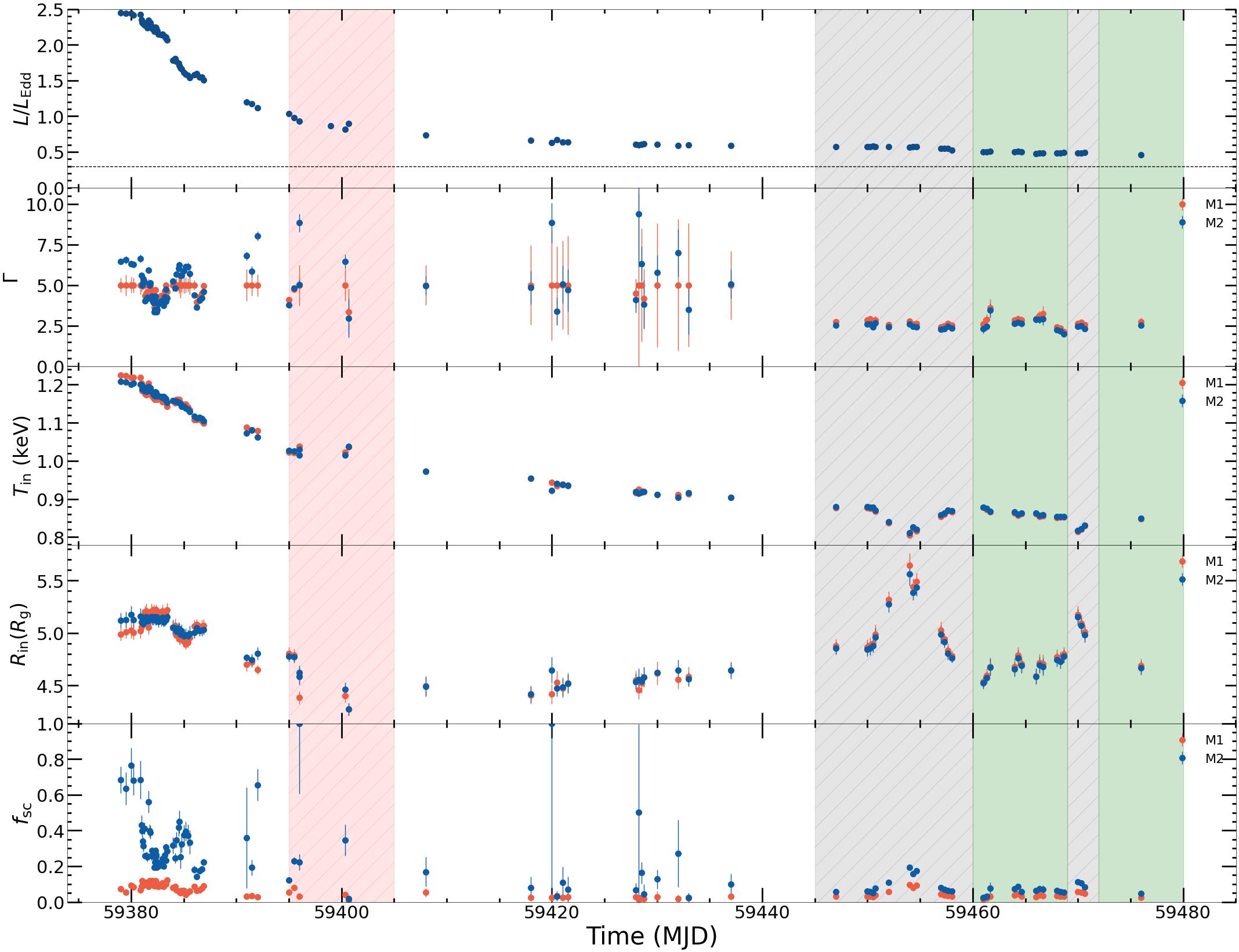}
    \caption{Spectral parameter evolution of 4U 1543--47 as observed by \textit{Insight}-HXMT, where $L/L_{\mathrm{Edd}}$ is the total disk luminosity (0.01 -- 30 keV) in Eddington units, with $L_{\mathrm{Edd}}$ = 1.2 $\times 10^{39}$ ergs/s and systematic uncertainties from the distance and BH mass accounted for; $f_{\mathrm{sc}}$ is the scattering fraction; $\Gamma$ is the photon power-law index of the Comptonization component; $T_{\mathrm{in}}$ is the inner disk temperature in units of keV; and $R_{\mathrm{in}}$ is the inner disk radii in units of gravitational radii $R_{\mathrm{g}}$. The shaded red regions denotes the epoch for state transition \citep{Jin2023}. The shaded green regions denote ideal epochs for spin extraction. Due to the similar luminosity values between M1 and M2 which appear indistinguishable, we present only the M2 luminosity values. The $\Gamma$ factors and $f_{\mathrm{sc}}$ between 59400 and 59440 cannot be constrained well due to the low photon count in the ME band for observations in this epoch.}
    \label{fig:fit7}
    
\end{figure*}

\begin{figure}[ht]
    \centering
    \includegraphics[width = 0.5\textwidth]{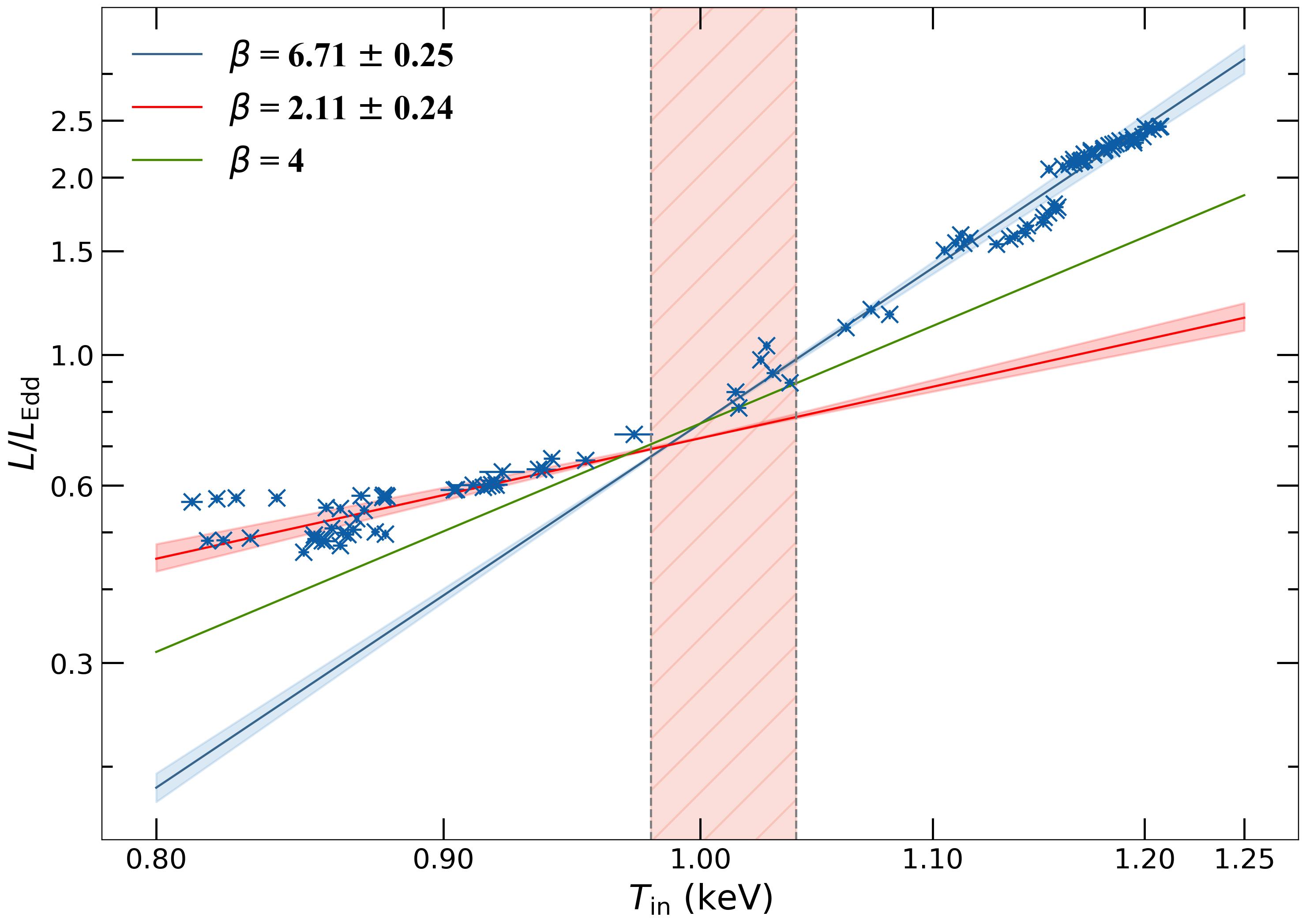}
    \caption{ Luminosity-temperature (L-T) relation (L $\propto T^{\beta}$) for super-Eddington observations  with M2 values, which yields a best-fit power value of $\beta = 6.71 \pm$ 0.25. The shaded red region denotes the disk state transition epoch found by \cite{Jin2023}. For observations after MJD 59405.0, the best-fit power value is $\beta = 2.11 \pm$ 0.24.  }
    \label{fig:LT}
\end{figure}

\begin{figure*}[h]
\centering
\begin{subfigure}[b]{.54\textwidth}

  \centering
  \includegraphics[width=.7\linewidth]{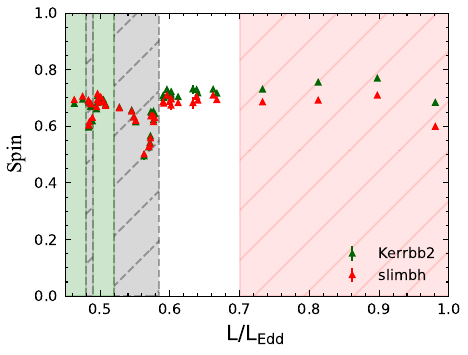}  
  \caption{$\alpha$ = 0.01}
  \label{fig:sub-first1}
\end{subfigure}
\hfill
\hspace{-5cm}
\begin{subfigure}[b]{.54\textwidth}
  \includegraphics[width=.7\linewidth]{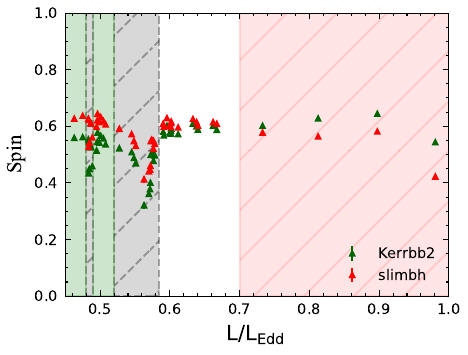}  
  \centering
  \caption{$\alpha$ = 0.1}
  \label{fig:sub-second2}
\end{subfigure}
\begin{subfigure}[b]{.55\textwidth}
  \centering
  \includegraphics[width=.7\linewidth]{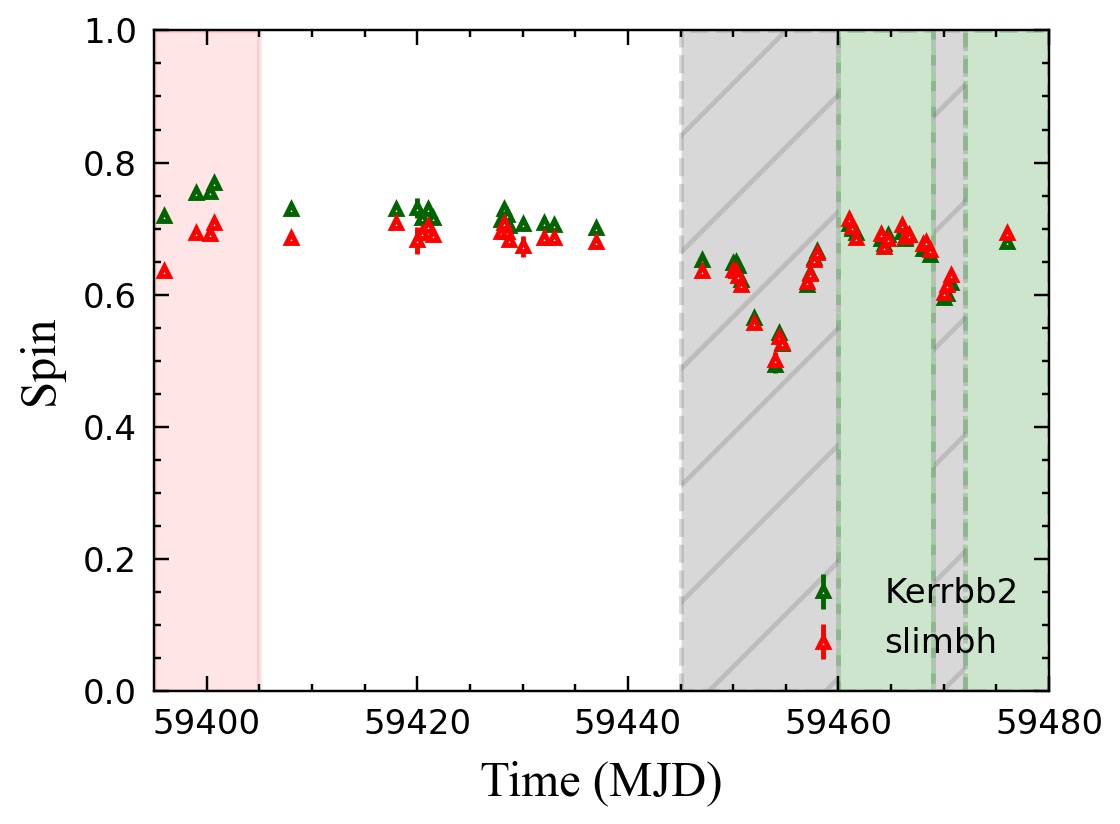}  
  \caption{$\alpha$ = 0.01}
  \label{fig:spintimelv}
\end{subfigure}
\hfill
\hspace{-4cm}
\begin{subfigure}[b]{.55\textwidth}
  \includegraphics[width=.7\linewidth]{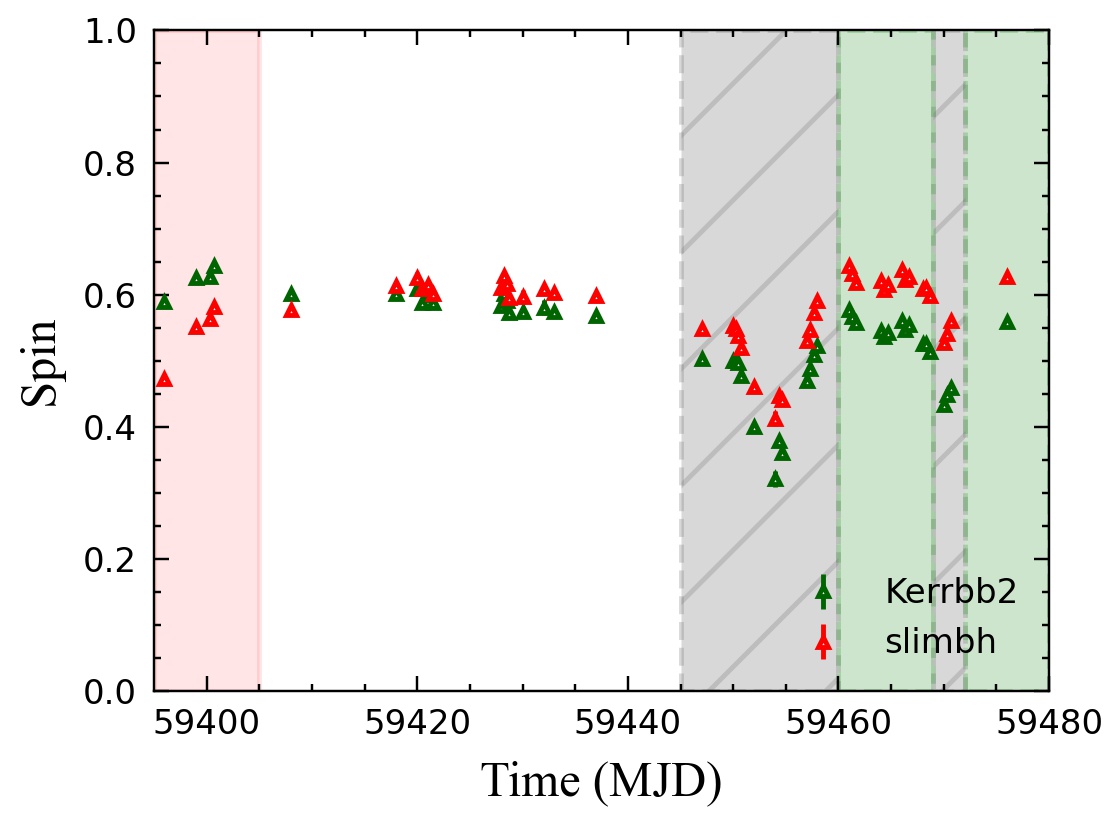}  
  \centering
  \caption{$\alpha$ = 0.1}
  \label{fig:spintimehv}
\end{subfigure}
\hfill
\hspace{-2.4cm}

\caption{Evolution of both \texttt{slimbh}  and \texttt{kerrbb2} spin estimates vs luminosity and time for all \textit{Insight}-HXMT observations below 1 $L_{\mathrm{Edd}}$. The dips around 0.55 $L_{\mathrm{Edd}}$ and 0.49 $L_{\mathrm{Edd}}$ correspond to the rise in inner disk radius and $f_{\mathrm{sc}}$  between MJD 59445--59460 and 59469--59472. Both \texttt{kerrbb2} and \texttt{slimbh} give stable spin values up to $\sim$0.9 $L_{\mathrm{Edd}}$.  }

\end{figure*}

\begin{figure*}[h]
    \centering
    \includegraphics[width = 0.53\textwidth]{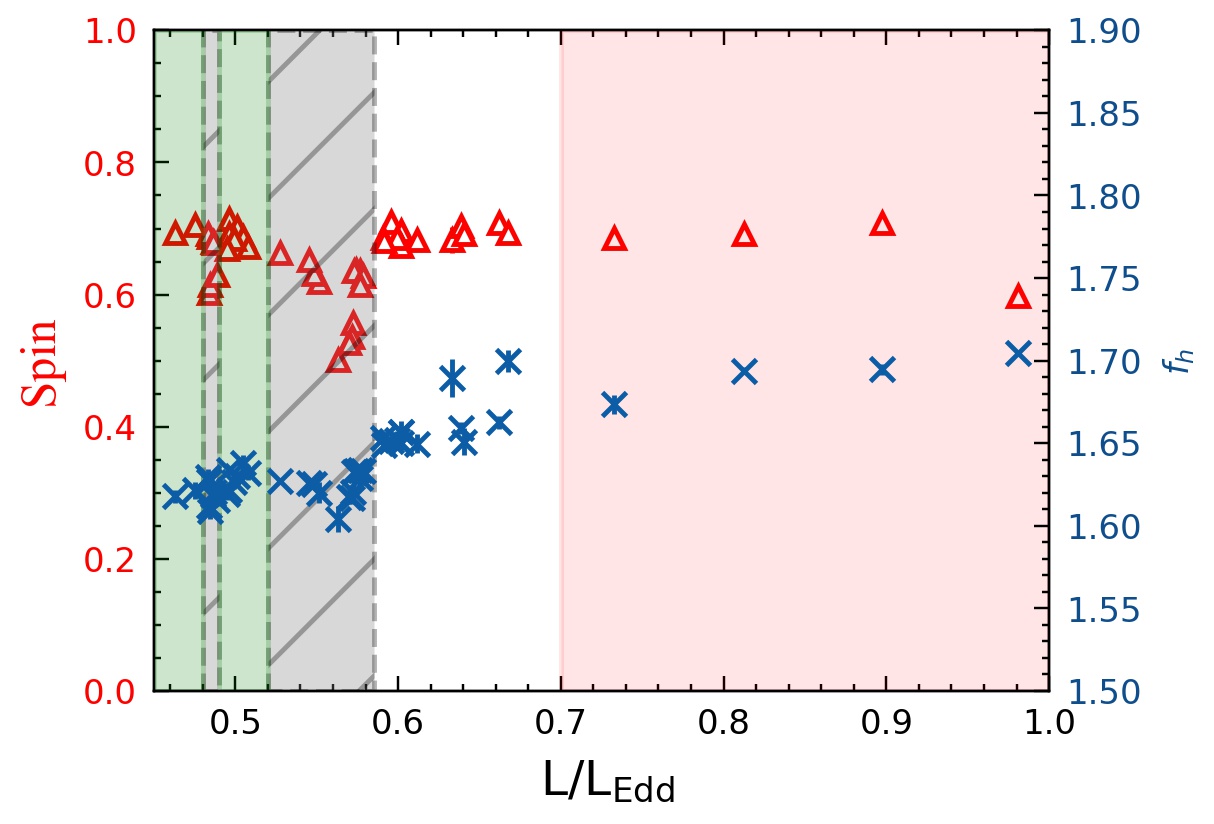}
    \caption{Luminosity dependence of the hardening factor $f_{\mathrm{h}}$ (blue) and spin $a_{*}$ (red) measured with slimbh ($\alpha = 0.01$) }
    \label{fig:fsc}
\end{figure*}

\begin{table*}[ht] \centering
\newcommand{\ra}[1]{\renewcommand{\arraystretch}{#1}}
\ra{1.4}

\begin{tabular}{rrrrrrr}
\multicolumn{7}{l}{\textbf{Table 4.} Fitting results with M2 on selected `golden' \textit{Insight}-HXMT data}\\[0.2cm] 
  \toprule 

 Observation \phantom{a} & $f_{\mathrm{sc}} $ \phantom{ab} & $\Gamma$          \phantom{ab} & $T_{\mathrm{in}}$ (keV) & $R_{\mathrm{in}}$ ($R_{g}$) &$l_{\mathrm{Edd}}$ &  $\chi^2 / $DOF          \\ 
 \hline

P030402603501  & $0.024^{+0.007}_{-0.007}$         &  $2.31^{+0.29}_{-0.29}$  &  $0.880^{+0.001}_{-0.001}$ &$4.57^{+0.06}_{-0.06}$&0.50& 1219/1048          \\
          P030402603502               & $0.032^{+0.008}_{-0.008}$          &  $2.46^{+0.28}_{-0.28}$        &  $0.880^{+0.002}_{-0.002}$&$4.68^{+0.10}_{-0.10}$&0.50& 1153/1048           \\
          P030402603503     & $0.078^{+0.029}_{-0.029}$         & $3.43^{+0.44}_{-0.44}$         &  $0.870^{+0.003}_{-0.003}$  &$4.65^{+0.06}_{-0.06}$&0.50& 1062/1048       \\
          P030402603601     & $0.074^{+0.010}_{-0.010}$         & $2.65^{+0.12}_{-0.12}$        &  $0.870^{+0.003}_{-0.003}$   &$4.76^{+0.07}_{-0.07}$&0.50& 1243/1048        \\
          P030402603602     & $0.087^{+0.015}_{-0.015}$         & $2.69^{+0.18}_{-0.18}$          &  $0.860^{+0.003}_{-0.003}$   &$4.69^{+0.07}_{-0.07}$&0.51& 1076/1048       \\
          P030402603603     & $0.059^{+0.012}_{-0.012}$         & $2.62^{+0.21}_{-0.21}$       &  $0.860^{+0.002}_{-0.002}$   &$4.59^{+0.07}_{-0.07}$&0.50& 1120/1048       \\
          P030402603701     & $0.065^{+0.014}_{-0.014}$         & $2.87^{+0.21}_{-0.21}$     &  $0.860^{+0.003}_{-0.003}$   &$4.69^{+0.07}_{-0.07}$&0.48& 1181/1048      \\
          P030402603702     & $0.075^{+0.019}_{-0.019}$         & $2.87^{+0.41}_{-0.41}$       & $0.860^{+0.003}_{-0.003}$  &$4.68^{+0.08}_{-0.08}$&0.48& 1045/1048         \\
          P030402603703     & $0.069^{+0.025}_{-0.025}$         & $2.91^{+0.11}_{-0.11}$      &  $0.850^{+0.003}_{-0.003}$  &$4.75^{+0.08}_{-0.08}$&0.48& 1075/1048        \\
          P030402603801     & $0.064^{+0.008}_{-0.008}$         & $2.23^{+0.12}_{-0.12}$       & $0.850^{+0.003}_{-0.003}$  &$4.75^{+0.07}_{-0.07}$&0.49&  1086/1048       \\
          P030402603802     & $0.056^{+0.007}_{-0.007}$         & $2.16^{+0.01}_{-0.01}$        &  $0.850^{+0.002}_{-0.002}$  &$4.73^{+0.07}_{-0.07}$&0.49&  1053/1048      \\
          P030402603803     & $0.052^{+0.005}_{-0.005}$         & $1.99^{+0.01}_{-0.01}$        &  $0.850^{+0.002}_{-0.002}$  &$4.78^{+0.06}_{-0.06}$&0.49&  1136/1048        \\
          P030402604101     & $0.047^{+0.01}_{-0.01}$         & $2.54^{+0.20}_{-0.20}$       &  $0.850^{+0.002}_{-0.002}$  &$4.66^{+0.07}_{-0.07}$&0.46&  1147/1048      \\
 \bottomrule
\end{tabular}
\caption*{\textbf{Notes.} $f_{\mathrm{sc}} $ is the scattering fraction, $\Gamma$  is the photon index, $T_{\mathrm{in}}$ is the inner disk temperature, $R_{\mathrm{in}}$ ($R_{g}$) is the inner disk radius in units of gravitational radii $R_{g}$, and   $l_{\mathrm{Edd}}$ is the disk luminosity in Eddington units.} 
\captionlistentry{Table}
\label{tab:fit9}
\end{table*}

We plot the evolution of the scattering fraction $f_{\mathrm{sc}}$, Comptonized photon index $\Gamma$, inner disk temperature $T_{\mathrm{in}}$, inner disk radius $R_{\mathrm{in}}$, and luminosity $L_{\mathrm{Edd}}$ obtained from both M1 and M2 in Figure \ref{fig:fit7}. The inner disk temperature and luminosity show a steep decrease in the initial phase of the outburst (MJD 59379.0 -- 59400.0), followed by a shallower decrease. The source starts with a high power-law index with $\Gamma \geq 5$ and scattering fraction ($f_{\mathrm{sc}}$ > 50 \% for M2); between MJD 59410.0--59440.0, the power-law index and scattering fraction cannot be constrained well due to the high background in ME. The source enters into a stable configuration between MJD 59460.0-59469.0 and MJD 59472.0--59480.0  with $ 2 < \Gamma < 3$ and $f_{\mathrm{sc}}$ < 10 \%, satisfying the criteria for spin extraction mentioned in section 3 (See green highlighted epoch, Figure \ref{fig:fit7}). Here, the inner disk radius  remains stable at around 4.6 $R_{\mathrm{g}}$. We therefore identify observations within these two epochs as those providing golden spectra from which to extract spin (See Table \ref{tab:fit9}). Between MJD 59418.0--59450.0,  the source also enters a period of a minimum, stable inner disk radius at around 4.5 $R_{\mathrm{g}}$. However, since $\Gamma$  and $f_{\mathrm{sc}}$ cannot be constrained well here due to the low photon count above 15 keV, these observations have been excluded.

BHB sources are expected to undergo several phases of exponential decay during transitions between different disk states within outbursts \citep{Eckersall2015}. \cite{Jin2023} fit the light curve and hardness-intensity diagram (HID) of the source with exponential curves, and identify two branches of the decay phase, pointing to a disk state transition from a thicker slim-disk to a thin-disk between MJD 59395.0 and MJD 59405.0. The source luminosity during this transition period incidentally crosses one Eddington. Motivated by this finding, we plot the disk luminosity and the inner-disk temperature (where $L$ and $T_{\mathrm{in}}$ follow $L$ $\propto T^{\beta}_{\mathrm{in}}$), and fit the curve for observations before and after the state transition (see Figure \ref{fig:LT}) identified by \cite{Jin2023}. We find a very steep curve before the state transition, with $\beta$ = 6.71 $\pm 0.25$, while after the state transition, the curve flattens with $\beta$ = 2.11 $\pm 0.24$.

For all observations < 1 $L_{\mathrm{Edd}}$, we compare the luminosity dependence of the spin from two relativistic accretion disk models by replacing {\ttfamily diskbb} with {\ttfamily kerrbb2} (M3) \citep{mcc2006} and {\ttfamily slimbh} (M4) \citep{Sadowski2011}. {\ttfamily kerrbb2} is based on the NT thin-disk solutions, and is a hybrid of {\ttfamily kerrbb} \citep{Li2005} and {\ttfamily BHSPEC} \citep{Davis_2006}.  The relativistic slim-disk model, {\ttfamily slimbh}, is a generalization of the standard thin-disk model, {\ttfamily kerrbb2}, taking into account the aforementioned deviations from thin-disk solutions presented in section 2, in particular, ray tracing which can be done from the disk photosphere rather than the equatorial plane. As is the case for {\ttfamily kerrbb2}, {\ttfamily slimbh} can account for the color correction factor, $f_{\mathrm{h}}$ (where $f_{\mathrm{h}}$ is the ratio of observed color temperature to effective temperature, $f_{\mathrm{h}} =T_{\rm{col}}/T_{\rm{Eff}}$),  with the aid of TLUSTY stellar atmospheres code spectra, which computes the vertical structure and radiation transfer in accretion disks, and can model the X-ray continuum up to the Eddington limit \citep{hubeny1995non}. The comparison of the spin between the two models was performed twice for two values of viscosity: 0.1 and 0.01 (Figure \ref{fig:sub-first1}, \ref{fig:sub-second2}). By doing so, we are able to better gauge (i) the effect of viscosity on spin measurements, and (ii) the disk structure by comparing the NT model and slim-disk models. $M$, $i$, and $D$ are fixed to their fiducial values. As mentioned, we use M3-M4, with $f_{\mathrm{val}}$ for \texttt{kerrbb2} set to -1, allowing the spectral hardening factor $f_{\mathrm{h}}$ to be interpolated over the table, and $f$ for \texttt{slimbh} is set to -1, allowing $f_{\mathrm{h}}$ to be obtained from TLUSTY spectra. Above 0.6 $L_{\mathrm{Edd}}$ for $\alpha$-viscosity =  0.01, and 0.7  $L_{\mathrm{Edd}}$ for $\alpha = $ 0.1, \texttt{kerrbb2} consistently estimates a higher spin value than \texttt{slimbh}, and the effect is more pronounced at higher luminosities. This is expected, because emission from within the ISCO is predicted by slim-disk models at higher accretion rates \citep{Straub2011}. There is excellent agreement below 0.6 $L_{\mathrm{Edd}}$ between the two models in the case of $\alpha$-viscosity = 0.01. We show the luminosity dependence of the hardening factor against the backdrop of the spin evolution obtained from \texttt{slimbh} in Figure \ref{fig:fsc}. The hardening factor was obtained by fixing the previously obtained spins for each respective observation, and setting the $f_{\rm{h}}$ parameter of \texttt{slimbh} free ($f_{\rm{h}}$ > 1) . The hardening factor starts around 1.7 at the state transition and gradually decreases to around 1.6 toward the end of \textit{Insight}-HXMT's observation epoch of the source. This is indeed consistent with previous findings; through a series of fits to synthetic spectra produced via TLUSTY code, \cite{davis2019spectral} found that $f_{\rm{h}}$ increases with inner disk temperature ($T_{\rm{Eff}}$) and accretion rates.

\subsection{Spin error analysis}

The bulk of the uncertainty on continuum fitting derives from the uncertainties in the BH mass, distance, and inclination angle. For our selected "golden" spectra, we perform monte carlo simulations of the mass, inclination angle, and distance following
the prescriptions of \cite{Gou_2011}. $M$ and $i$ can be decoupled with the aid of the mass function, $f(m) = \frac{M^{3}\sin{i}^{3}}{(M+M_{opt})^2} $ = 0.25 ± 0.01 \citep{par2004}, where $M_{\mathrm{opt}}$ is the mass
of the optical companion ($2.7 \pm 1M_{\odot}$,  \citealt{par2004}). To this end, for each spectrum, we (1) generated 2000 parameter sets of $f(m)$, $i$, $D$, and $M_{\mathrm{opt}}$, (2) solved for the source mass M for a set of $f(m)$, $i$, $D$, and $M_{\mathrm{opt}}$, (3) generated a look-up table for each set, and
(4) refitted the spectrum with models M4 (\textit{Inight}-HXMT)and M5 (\textit{NICER}) to obtain the distribution of $a_{*}$. This process was repeated for both viscosity parameters of 0.1 and 0.01.

\subsection{\textit{Insight}-HXMT}

We identify 13 sub-exposures observed during the golden epoch identified in Section 4.1.  We fitted the selected \textit{Insight}-HXMT datasets with M4 in the 2-30 keV band, with a systematic error of 1\% for LE, again with \cite{wil2000} abundances and \cite{ver1996} cross sections. The switch for limb darkening is off (\textit{lflag} = -1), while (\textit{rflag} = 1) allowing raytracing to be done from the photosphere and taking into account vertical disk thickness.  The normalization is fixed to 1. $f_{\rm h}$ was set to -1 allowing the spectral hardening factor to be interpolated from TLUSTY.   The summed spin histogram distributions for our chosen golden \textit{Insight}-HXMT  spectra are shown in Figures \ref{fig:sub-first} and \ref{fig:sub-spin2}. Combined, we estimate a spin value of $\alpha_{*} = 0.64^{+0.14}_{-0.24}$ and $\alpha_{*} = 0.56^{+0.17}_{-0.27}$ for $\alpha = $ 0.01 and 0.1, respectively. The spectra are well fitted with an average reduced $\chi^2$ statistic of 0.99 and 1.04  for $\alpha$-viscosity =  0.01, 0.1, respectively (see table \ref{tab:testspinresult}).

\subsection{NICER}

\begin{figure*}[h!]
\begin{subfigure}{.61\textwidth}

  \centering
  \includegraphics[width=.73\linewidth]{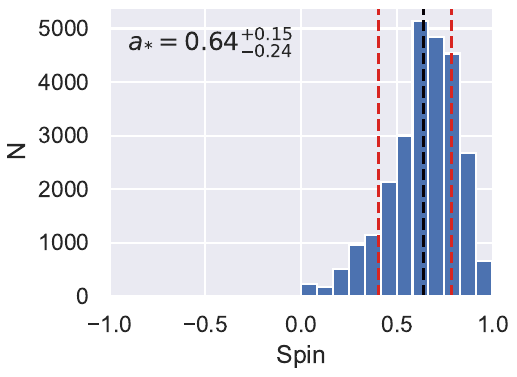}  
  \caption{\textit{Insight}-HXMT,  $\alpha$-viscosity = 0.01}
  \label{fig:sub-first}
\end{subfigure}
\hfill
\hspace{-3.0cm}
\begin{subfigure}{.61\textwidth}
  \centering
  \includegraphics[width=.73\linewidth]{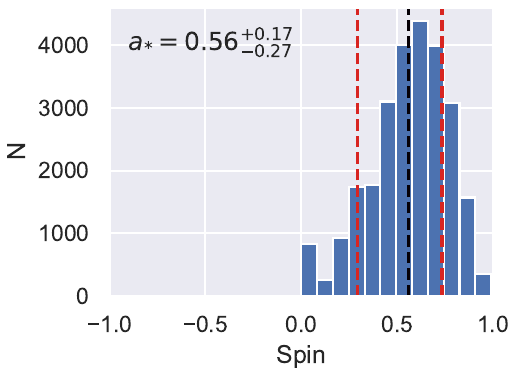}  
  \caption{\textit{Insight}-HXMT,  $\alpha$-viscosity = 0.1}
  \label{fig:sub-spin2}
\end{subfigure}
\begin{subfigure}{.61\textwidth}

  \centering
  \includegraphics[width=.73\linewidth]{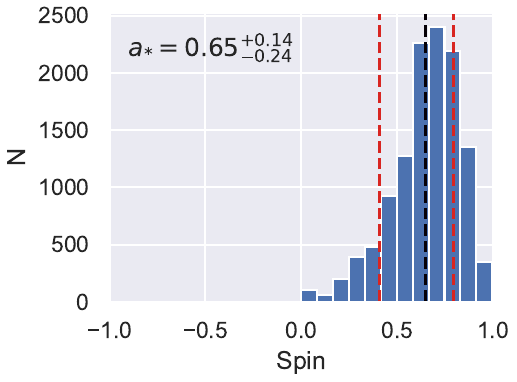}  
  \caption{\textit{NICER},  $\alpha$-viscosity = 0.01}
  \label{fig:sub-spin3}
\end{subfigure}
\hfill
\hspace{-3.0cm}
\begin{subfigure}{.61\textwidth}
  \centering
  \includegraphics[width=.73\linewidth]{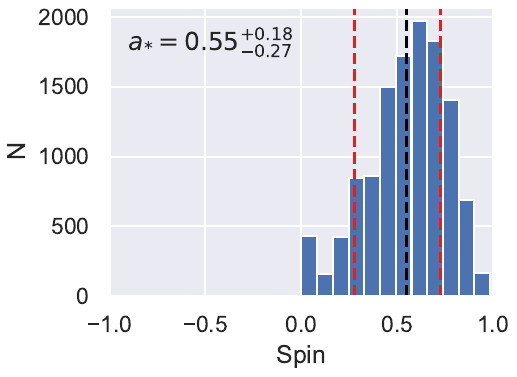}  
  \caption{\textit{NICER},  $\alpha$-viscosity = 0.1}
  \label{fig:sub-second}
\end{subfigure}
\caption{Summed spin distributions for \textit{Insight}-HXMT and \textit{NICER} from MC analysis, with $\alpha$ viscosity = 0.01 (\ref{fig:sub-first},  \ref{fig:sub-spin3}) and $\alpha$ viscosity = 0.1 (\ref{fig:sub-spin2},  \ref{fig:sub-second}). Dashed lines denote 1$\sigma$ errors. }
\label{fig:spintotal3}

\end{figure*}

 We identify six \textit{NICER} observations performed between MJD 59460.0 and 59469.0, and between 59472.0 and 59480.0. No GTIs survived between MJD 59473.0 and 59480.0 because of abnormal background conditions, and so six NICER observations were excluded \footnote{https://heasarc.gsfc.nasa.gov/lheasoft/ftools/headas/nibackgen3C50.html}.  We consider the energy range between 2 and 9 keV, this time using M5, which does not have a Comptonized component because higher Comptonized energies are not used. We applied the same abundances, column density, and systematic as those used for \textit{insight}-HXMT. The smeared edge width was fixed to 7 keV, and an additional edge component was added (\texttt{edge}) and fixed to 2.0 keV to account for the Au edge. Combined, we estimate a spin value of $\alpha_{*} = 0.55^{+0.18}_{-0.27}$ for $\alpha = $ 0.1, and $\alpha_{*} = 0.65^{+0.14}_{-0.24}$ for $\alpha = $ 0.01 (See Figure \ref{fig:sub-spin3} and \ref{fig:sub-second}). The spectra are well fitted with an average reduced $\chi^2$ statistic of 0.99 and 1.03 for $\alpha$-viscosity =  0.01 and  0.1, respectively   (see table \ref{tab:testspinresult}).

\begin{table*}[ht!]
\centering
\def\arraystretch{1.3}
\begin{tabular}{l c c c c}
\multicolumn{5}{l}{\textbf{Table 5.} Spin estimates of all `golden' observations.}\\[0.2cm]
\toprule
& \multicolumn{2}{c}{$\alpha$-viscosity = 0.01} & \multicolumn{2}{c}{$\alpha$-viscosity = 0.1} \\ 
\cmidrule(lr){2-3} \cmidrule(lr){4-5}
&  $a_{*}$ &  $\chi^2 / $DOF &  $a_{*}$ & $\chi^2 / $DOF \\ 
\midrule
\textit{Insight}-HXMT & & & & \\
P030402603501 &  $0.66^{+0.14}_{-0.23}$ & 1102/1048 & $0.58^{+0.17}_{-0.27}$ & 1180/1048 \\
P030402603502 & $0.65^{+0.15}_{-0.23}$ & 1074/1048 &  $0.57^{+0.17}_{-0.26}$ & 1160/1048 \\
P030402603503 & $0.64^{+0.15}_{-0.24}$& 960/1048 &  $0.56^{+0.17}_{-0.27}$& 1010/1048\\
P030402603601 &  $0.64^{+0.11}_{-0.18}$& 1181/1048 &  $0.56^{+0.17}_{-0.26}$& 1221/1048\\ 
P030402603602 & $0.63^{+0.15}_{-0.24}$& 950/1048&  $0.55^{+0.17}_{-0.26}$& 993/1048\\
P030402603603 &  $0.63^{+0.15}_{-0.24}$& 1045/1048&  $0.55^{+0.18}_{-0.26}$& 1101/1048\\
P030402603701 &  $0.63^{+0.12}_{-0.18}$& 1156/1048&  $0.58^{+0.17}_{-0.26}$& 1188/1048\\
P030402603702 &  $0.64^{+0.15}_{-0.23}$& 947/1048&  $0.56^{+0.17}_{-0.26}$& 987/1048\\
P030402603703 &  $0.65^{+0.15}_{-0.23}$& 977/1048&  $0.57^{+0.17}_{-0.26}$& 1005/1048\\
P030402603801 &  $0.63^{+0.15}_{-0.24}$& 992/1048&  $0.55^{+0.18}_{-0.26}$& 1033/1048\\
P030402603802 &  $0.63^{+0.15}_{-0.24}$& 993/1048&  $0.55^{+0.18}_{-0.26}$& 1037/1048\\
P030402603803 &  $0.61^{+0.15}_{-0.25}$& 1051/1048&  $0.53^{+0.18}_{-0.26}$& 1113/1048\\
P030402604101 &  $0.64^{+0.15}_{-0.23}$& 1077/1048&  $0.56^{+0.17}_{-0.26}$& 1121/1048\\
\textit{NICER} & & & & \\
4202230162 &  $0.61^{+0.15}_{-0.23}$ & 630/699 &  $0.50^{+0.20}_{-0.25}$& 698/699\\
4202230165 &  $0.64^{+0.14}_{-0.23}$ & 625/699 &  $0.54^{+0.18}_{-0.26}$& 670/699\\
4202230166 &  $0.65^{+0.14}_{-0.23}$ & 584/699 &  $0.54^{+0.18}_{-0.26}$& 647/699\\
4202230167 &  $0.63^{+0.15}_{-0.25}$ & 604/699 & $0.52^{+0.19}_{-0.25}$& 662/699\\
4202230168 &  $0.69^{+0.13}_{-0.23}$ & 631/699 &  $0.60^{+0.16}_{-0.28}$& 656/699\\
4202230172 &  $0.68^{+0.13}_{-0.21}$ & 1094/699 &  $0.59^{+0.16}_{-0.27}$& 1015/699\\
\bottomrule
\end{tabular}
\caption*{\textbf{Notes.} The energy range for \textit{Insight}-HXMT is 2--8 keV (LE) \& 10--30 keV (ME). The energy range for \textit{NICER} is between 2--9 keV. Energies below 2 keV for both \textit{Insight}-HXMT \& \textit{NICER} were excluded due to poor calibration.}
\captionlistentry{Table}
\label{tab:testspinresult}
\end{table*}

\section{Discussion}
\subsection{Disk evolution}

Our study of the spectral evolution of the source shows that the source undergoes a disk transition from a slim disk to a standard thin-disk around 1 $L_{\mathrm{Edd}}$, and more detailed studies are ongoing on this specific subject. This is suggested by the evolution of the L-T relationship (Figure \ref{fig:LT}), where the source starts out with a steep power law with $L \propto T_{\mathrm{in}}^{6.71}$ before MJD 59395.0, and shallows to $L \propto T_{\mathrm{in}}^{2.11}$ after MJD 59405.0. It is generally expected that in the case of a fixed emitting area, the luminosity would scale with temperature  with $L\propto T^{4}_{\mathrm{Eff}}$. A shallower relationship may indicate an unstable inner disk radius, and this is indeed the case, because the power-law fitting includes  epochs where the inner disk recedes from the ISCO (MJD 59445.0 -- 59460.0 and MJD 59469.0 -- 59472.0; see epochs shown in grey in Figure \ref{fig:fit7}). Moreover, as $f_{\rm{h}}$ decreases with decreasing temperature, and  $T_{\mathrm{in}} = f_{\rm{h}}T_{\rm{Eff}}$, $f_{\rm{h}}$ > 1, the modified temperature would further shallow the L-T relation. In addition, advection may still play a role in these moderate luminosities below 1 $L_{\mathrm{Edd}}$. Work by \cite{watarai2000galactic} showed that in the slim-disk regime, $T_{\rm{in}} \propto r^{-1/2}$, suggesting that slim-disk models should render $\beta \sim 2$. Therefore, $\beta \sim 2$ is not surprising. More intriguing is the very steep relation before the disk state transition ($L/L_{\rm{Edd}}$ > 1). There are two possible explanations for this at these super-Eddington observations: either the disk temperature profile deviates from the prediction of Shakura and Sunyaev that $T(r) \propto r^{-3/4}$ in unforeseen ways, or a model decomposition issue is the culprit, given that many of these observations have a larger fraction of emission in the harder component.

The ``evolution'' of the spin over time also seems to indicate a state transition. Although \cite{Straub2011} find a clear luminosity dependence of the spin measured by both slim and NT models, we find both models to be relatively stable for both values of $\alpha$ viscosity. This gives further credence to the state transition hypothesis around 1 $L_{\mathrm{Edd}}$. Furthermore, the fact that $f_{\rm{h}}$ decreases with decreasing $T_{\mathrm{in}}$ while obtaining a constant spin suggests a softening of the spectrum with decreasing luminosity, which is consistent with model predictions \citep{davis2019spectral}.

In all previous CF spin measurements, the thin-disk criterion of < 0.3 $L_{\mathrm{Edd}}$ was strictly  employed when using continuum fitting for spin measurements of BHBs. Our results suggest the possibility of recovering thin-disk solutions at even higher luminosities. However, we have chosen to remain conservative and use the slim-disk model in our spin measurements, as all our selected 'golden' spectra have luminosities greater than > 0.45 $L_{\mathrm{Edd}}$. Between  0.52 $L_{\mathrm{Edd}}$ -- 0.58 $L_{\mathrm{Edd}}$, and 0.48 $L_{\mathrm{Edd}}$ -- 0.49 $L_{\mathrm{Edd}}$, there is a sharp drop in spin. The time of these observations corresponds to the rise in the inner disk radius between MJD 59445.0 -- 59460.0 \& MJD 59469.0 -- 59472.0, a rise in the scattering fraction $f_{\mathrm{sc}}$, along with a dip in the inner disk temperature $T_{\mathrm{in}}$ (see epochs marked in gray in Figures \ref{fig:fit7}, \ref{fig:fsc}). A sharp hardening in the former epoch is found in the HR in Figure \ref{fig:lc_hr}, and hard flares have been detected by ME, HE, and AstroSAT \citep{prabhakar2023wideband}, suggesting that the source enters an intermediate hard state before transitioning back to a soft state, and this would manifest itself as lower spin estimates. This epoch may also correspond to the presence of jet emissions, because jets driven by magnetic fields around accretion disks will use up part of the accretion power, which will result in a reduction of the disk temperature \citep{blandford1982hydromagnetic, li2014thermal, li2009global}. In the meantime, a receding disk is also observed, suggesting that a jet base may occur in the inner part of the accretion disk pushing material out, hence the increased inferred inner disk radius \citep{ferreira2006unified, marcel2022unified}. We  exclude these epochs from our final spin estimate. Between MJD 59460.0 and 59469.0, and between MJD 59472.0 and 59480.0 (green epochs, Figure \ref{fig:fit7}), the constant, low inner disk radius coupled with photon index in the typical soft range 2 < $\Gamma$ < 3 and a low scattering fraction $f_{\mathrm{sc}}$ < 10\% make our implicit assumption that $r_{\mathrm{in}} = r_{\mathrm{ISCO}}$ most robust. 

\subsection{$\alpha$-viscosity}

Although the underlying mechanism behind angular momentum transport within accretion disks remain unknown, it is generally accepted that the turbulence arising from magneto-rotational instability is the main driver of accretion. Typically, continuum fitting is performed with the dimensionless viscosity parameter fixed at 0.1. Indeed, strong observational evidence points toward a value of between 0.1 and 0.4 to describe fully ionized, thin-disks
(see, e.g., \cite{dubus2001disc}, \cite{cannizzo2001v}, \cite{schreiber2003delays}). However, numerical simulations tend to estimate $\alpha$ an order of magnitude lower than what is suggested by observations ($\alpha \leq 0.02$, see \cite{King} and references therein). Although measurements of the $\alpha$ viscosity tend to be less reliable in the case of partially ionized disks, estimates are consistently finding values of $\alpha$ an order of magnitude smaller ($\alpha \sim 0.01$) than those for fully ionized disks \citep{Martin_2019}.

The lower spin measurements from assuming a higher $\alpha$-viscosity of 0.1 is not surprising; higher values of $\alpha$-viscosity yield lower disk densities, which would in turn yield harder spectra (higher $f_{\rm{h}}$), and therefore lower spin. This is because the ratio of absorption opacity to scattering opacity decreases with decreasing disk density \citep{davis2019spectral}. Although all former thermal spin measurements of 4U 1543--57 fixed $\alpha$ to 0.1, we take the $\alpha$ viscosity of 0.01 to be a more accurate dimensionless representation of angular momentum transport in our selected spectra for several reasons. Firstly, although iron emission lines for the source are prominent and wide during the start of the outburst, they are largely diminished after the state transition between MJD 59395.0 and 59405.0; the spectral analysis by \citep{Jin2023} finds somewhat diminished ionization $\xi$ and a diminished contribution from the  reflection component after MJD 59400.0. Hence, we take the disk to be partially ionized for the `golden' observations. In addition, there is excellent agreement between \texttt{kerrbb2} and \texttt{slimbh} for $l_{\mathrm{Disk}} < 0.6 L_{\mathrm{Edd}}$ in the case of $\alpha$ = 0.01, while the difference between the models is around $\approx$ 0.07 for $\alpha$ = 0.01.

\subsection{System parameter accuracy}

Indeed, the bulk of uncertainty on the spin measurements arises from uncertainties in $M$, $i$, and $D$ \citep{Gou2011}. Accurate measurements of the spin therefore entail accurate and precise measurements of these orbital parameters. The reflection studies of 4U 1543--47 performed by \cite{mil2009} and \cite{Dong2020}  both constrained the inclination to $\sim$ 10° higher than the value found by \cite{par2004} and adopted by us. It is unlikely that this is an indication of a misalignment between the orbital inclination angle ($\sim$ 21°) and the inclination angle of the inner disk, because accretion would have already torqued the system into alignment, given that the timescale for such an alignment is on the order of $10^{6}$ -- $10^{8}$ years \citep{mar2008}. Regarding distance, the second data release (DR2) by \textit{Gaia} gives a distance of $8.60^{+3.84}_{-2.46}$ kpc inferred from a Bayesian estimation of the parallax,  which is consistent with the fiducial distance of $7.5 \pm 1.0 $ kpc found from photometry \citep{Ghandi_2019}. When fixing the distance to 8.6 kpc, the final spin estimate from MC analysis decreases to $0.47^{+0.16}_{-0.22}$ for $\alpha$ = 0.01, which is consistent with the error range in the case of $D$ = $7.5 \pm 1.0 $ kpc. However, considering that this \textit{Gaia} distance was obtained from a Bayesian inference of the parallax - which may not be accurate at long distances \citep{astraatmadja2016estimating} - as well the fact that we explore the possibility of measuring the spin at high luminosities to compare our results to previous spin measurements, we still use the same distance of $D$ = $7.5 \pm 0.5 $ kpc by \cite{jonker2004distances} in reporting a final spin result.

\section{CONCLUSIONS}
We carefully measured the spectral evolution of 4U 1543--47 throughout its 2021 outburst with data provided by \textit{Insight}-HXMT, and compared the performance between two accretion disk models, {\ttfamily slimbh} and { \ttfamily kerrbb2}. The consistent and constant spin measurements over luminosities extending up to the Eddington limit, coupled with the two different regimes of the L--T relationship point to a disk state transition occurring around $\sim$ 1 $L/L_{\rm{Edd}}$.  Our results indicate that thin-disk solutions can be recovered at luminosities higher than the typical thin-disk selection criterion of 0.3 $L_{\mathrm{Edd}}$.
We have rigorously filtered observations for epochs when the source is most likely truncated at the ISCO, and measured the spin with both \textit{Insight}-HXMT and \textit{NICER}. Although all past measurements of spin for this source have assumed a viscosity parameter of 0.1, we adopt a lower  $\alpha$-viscosity parameter of 0.01 given evidence of a partially ionized disk and the consistency between the NT and slim-disk model, and report a final spin of $a_{*} = 0.65^{+0.14}_{-0.24}$ when combining \textit{Insight}-HXMT and \textit{NICER}, which is in good agreement with the reflection results obtained by \cite{Dong2020}, and is also broadly consistent with the results of \cite{sha2006}. 

\section{Acknowledgements}

This work is supported by the National Key R\&D Program of China (2021YFA0718500). This work made use of the data from the \textit{Insight}-HXMT mission, a project funded by China National Space Administration (CNSA) and the Chinese Academy of Sciences (CAS), as well as data obtained through the High Energy Astrophysics Science Archive Research Center Online Service, provided by the NASA/Goddard Space Flight Center. Q.B. acknowledges support by the National Natural Science Foundation of China (NSFC) under grants U1938102 and U1938107. A.V. acknowledges support from the Bundesministerium f{\"u}r Wirtschaft und Energie through Deutsches Zentrum f{\"u}r Luft-und Raumfahrt (DLR) under the grant number 50 OR 1917.


\bibliographystyle{aa}
\bibliography{aa}

\end{document}